# Mass diffusion in multilayer systems:
# an electrical analogue modelling approach

*Pawel Rochowski[1*], Giuseppe Pontrelli[2]*

[1]Institute of Experimental Physics, Faculty of Mathematics, Physics and Informatics, University of Gdansk, Wita Stwosza 57, 80-308 Gdansk, Poland
[2]Istituto per le Applicazioni del Calcolo, CNR, Italy
*Corresponding author: pawel.rochowski@ug.edu.pl, pawel.rochowski@gmail.com

**Abstract**

We develop a lumped parameter model to describe and predict the mass release of (absorption from) an arbitrary shaped body of any dimension in a large environment. Through the one-to-one analogy between diffusion-dominated mass transfer systems and electrical circuits we provide exact solutions in terms of averaged concentrations and mass released. An estimate of the release and characteristic time is also given. The proposed electrical analogue approach allows the definition of a time constant and lets an easy extension to a multi-layer and multi-phase cases. The simulation results are compared with those obtained from the solution of the corresponding analytical and numerical solutions, showing a good agreement.

1. Introduction

Diffusion occurs in several physical process, and involves all scales, from transport of chemicals in the cell to the dispersal of pollutants in sea water. Diffusive processes occur both in nature and in engineering applications and also underlie other phenomena such as solute-solvent interaction, solvation and turbulence (Cussler, 1997). In most circumstances and in simpler geometries, diffusing systems can be described by linear equation and often an analytical solution is available (Crank, 1975). When other phenomena are taken into account or the nature of diffusion involves more layers or phases, the complexity of the problem may become very difficult and excessively time consuming. In some cases, the use of reduced-order model for diffusion-controlled release systems has been proposed (Carr, 2021). Because of these difficulties, scientists and engineers concerned with the design of such systems have, with increasing frequency, made use of analogies. Thus an electrical system may be analogous to a mechanical, thermal, or other system provided there is a "likeness" between the two systems.

Usually such a likeness is not only of appearance of the mathematical formulation, but a resemblance of physical behavior. Thus, if it can be shown that the mathematical descriptions of the behavior of two different systems are similar , then it is usually possible to use one system as a means for studying the behavior of the other. Electrical circuit analogy offers an effective method for the description, and sometimes for the prediction, of various heat and mass transfer processes (McCann and Wilts, 1949; Paschkis and Baker, 1942; Paschkis and Heisler, 1944).

The ease with which electrical circuits can be assembled and their behavior have made them particularly useful as an analog systems (Paschkis and Heisler, 1944). Analogs have long been used by engineers, and hundreds of publications have appeared on the subject of electric analog technique (McCann and Wilts, 1949; Paschkis and Baker, 1942; Paschkis and Heisler, 1944). One of the better introductions to the great versatility of analog techniques in the solution of engineering problems is ref (Sorcka, 1954). Most work concerns of the extensive use of electrical circuits for analysis of transient heat-flow phenomena (Robertson and Gross, 1958). Lumped parameter models based on electric analogues for the human circulatory system dates back from 60's (Westerhof et al., 1969), and recently this analogy has been used to solve more complicated conduction problems, and for the performance analysis of heat exchange networks (Chen et al., 2015).

However, to the best of our knowledge, the analogy of mass transfer in releasing/absorbing conditions with electrical circuits has not been fully exploited. This would allow the characteristic features of even complex systems, including the estimation of the release time. In this article we fill the gap and we use the thermal-electrical analogy to propose a new sort of models to give deeper information about the intrinsic elements of the system, comprising the case of multi-layer and multi-component system, that have the counterpart in electrical resistances connected in parallel or in series. When the spatial distribution of concentration is not important, but one is more interested to the global behavior of a diffusive system and the characteristic time, it is common to use electric analogue where two or more compartments are connected. In addition, the state space description can be used to design advanced controllers with better performance. The method has demonstrated to recover the classical analytical solution of mass diffusion system (Crank, 1975) and can be adapted to multi-layer /multi-phase diffusive systems.

We develop an electrical analogue model for describing the transient mass diffusion with forward /reverse (donor/acceptor) transfer rates. The novel idea presented in this article is to build a compartmental geometry-independent model based on the one-by-one analogy that exists between mass releasing systems and electrical circuits. This analogy, that is shared with heat transfer problems, has been already mentioned in some heat transfer articles (Chen et al., 2015; Robertson and Gross, 1958), and it is basically sustained in the similarity of governing equations. The simplicity of the method is in contrast with analytical and numerical solutions that, in general, are geometry-dependent and become increasingly cumbersome and computationally inefficient in complex geometries.

The remaining sections of this paper are organized as follows. In the next section, we present the general concept of the 1-to-1 analogy between electrical and diffusive systems. In section 3, based on this analogy, we derive the solution of the problem, either in the general and special cases. The approach is extended to the case of multi-layer and composite media in Section 4 and to the two-phase system in sect. 5. Finally in sect. 6 we show how our solution can be viewed in terms of classical solutions and reduce to well-known series of exponential type for special geometries. More validation is made with other exact or approximated solutions, showing a good agreement.

## 2. Electric circuit analogue for diffusion-driven mass transfer

Schematically, a diffusion-dominated mass transfer system (MTS for short) relies on a concentration gradient between two points (or regions, or compartments), say $A$ and $B$. The direction of the transfer (*absorption or release*) strictly depends on the relative initial concentrations[1] $c_A^0$ and $c_B^0$. If $c_A^0 > c_B^0$, then $A$ is considered as a mass source (donor) and $B$ is considered as a mass recipient (acceptor) (*absorption*), the reverse when $c_A^0 < c_B^0$ (*release*)(fig.1). Regardless of the specific geometry, the mass transport driving force is proportional to the concentration difference that generates a mass flux and applies as long as an equilibrium ($c_A = c_B$) is reached. As such, this process has a close similarity with the electric circuits (shortly EC): an electric potential gradient between two points of a conductor induces a charge transfer, that gives rise to an electrical flux (current). In this section we aim at describing this analogy in a systematic way and build a space-free compartmental framework (lumped-parameter model) for diffusing MTS.

---

[1] Concentration is meant here as volume averaged concentration (c=M/V)

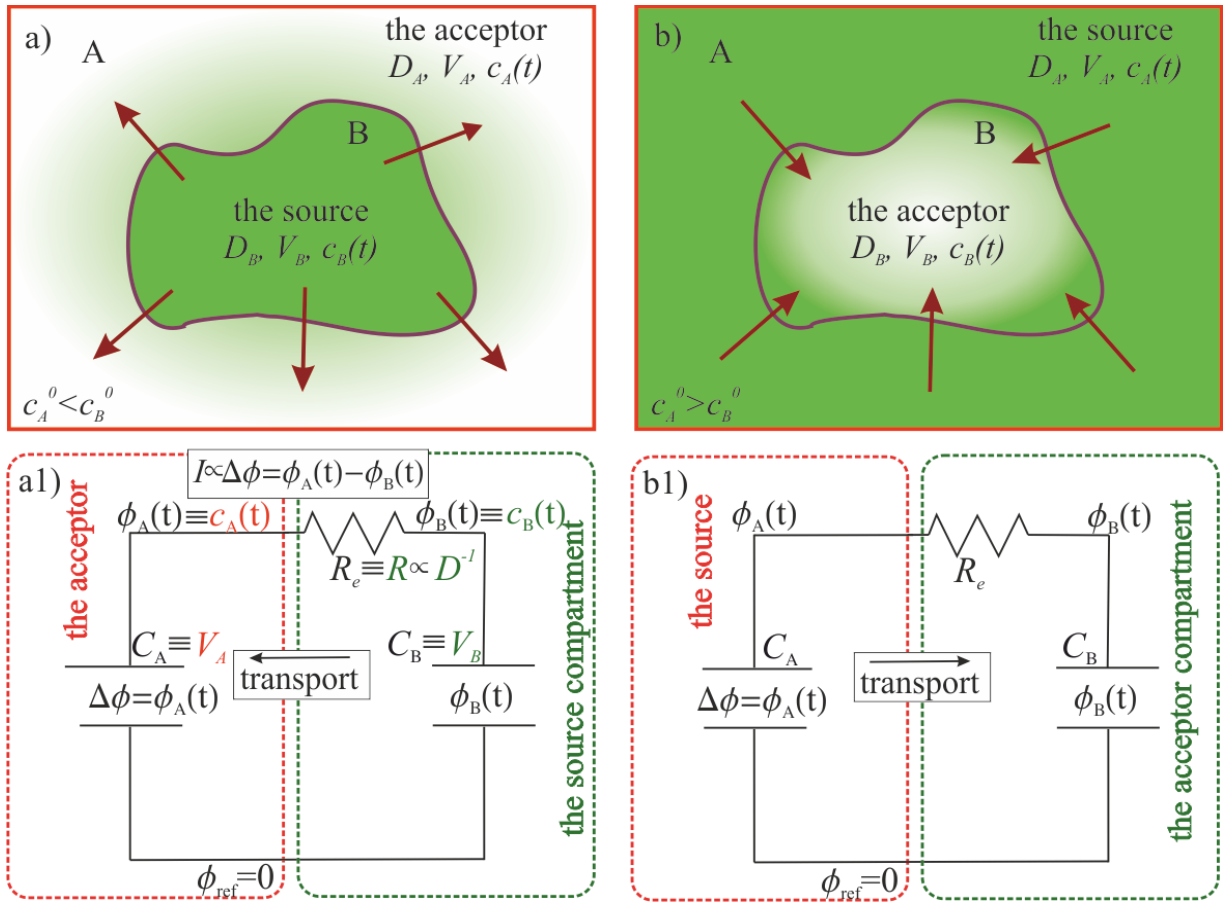

Fig 1. Analogies between the MTS and EC, where *A* represents a surroundings in which a releasing (a) or an absorbing (b) body *B* resides. Figs (a1) and (b1) represent the correspondent *RC* circuit diagrams.

To start with, let us consider a MTS made up of two contiguous bodies, here also referred to as compartments (to fix ideas, the body B is immersed in the environment A, with A and B of arbitrary geometries), see Fig.1a -1b. The two compartments are fully characterized by volumes ($V_A$, $V_B$) and diffusivities ($D_A$, $D_B$). In some circumstances, and in analogy with electrical conductors, the compartmental volume can be considered as a product of its exposed surface *a* by a characteristic length *l* ($V \approx a\, l$), see fig. 2. The characteristic length and the surface depend on the specific geometrical shapes of the objects; for example in the simplest cases of a plane slab or a sphere, *l* is directly related to the slab thickness or the sphere radius, respectively, while *a* is related to the external body surface. It should be noted, however, that the EC analogue model does not incorporate any assumption on the source and recipient shapes, dimensional geometry and coordinate system, with the exception of a perfect contact between the bodies A and B[2]. Typical circuits serving as surrogates for the diffusion-driven mass absorption/release processes are shown in fig.1.

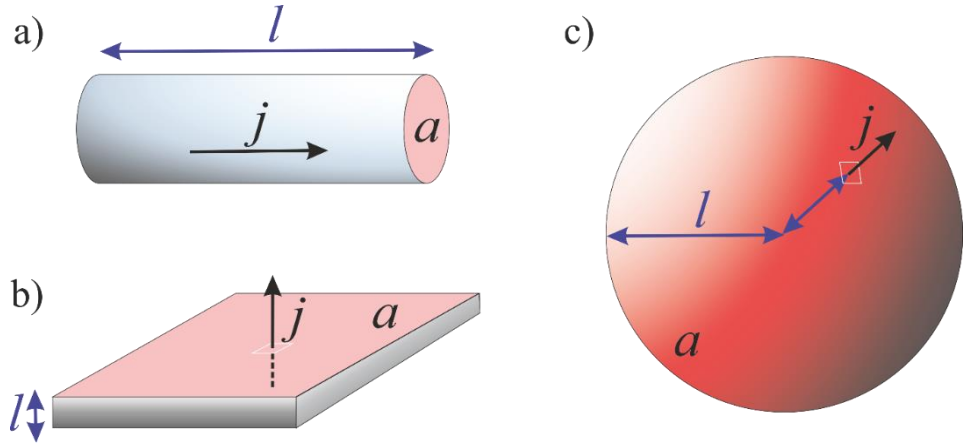

Fig.2: (a) Schematic representation of a portion of an EC: in correspondence of potential difference at the ends of a "conductor body" of length $l$ and cross section $a$, a flux $j$ (proportional to a current) is obtained. A similar mechanism occurs in the MTS due to a concentration difference (planar (case b), and spherical (case c) geometries).

To further explore analogies between the charge ($e$) and mass ($m$) transport schemes, first we refer to the 1D Ohm's and Fick's I laws, which give the electric and mass fluxes as:

$$j_q = -\kappa_e \nabla \phi, \qquad (2.1a)$$

$$j_m = -D \nabla c. \qquad (2.1b)$$

By comparing Eqs 2.1a and 2.1b, a direct correspondence between the electrical conductivity and the diffusion coefficient ($\kappa_e \equiv D$) and between the electric field potential and concentration ($\phi \equiv c$) emerge.

In an EC, by setting a zero value for the reference potential ($\phi_{ref} = 0$), the measured voltage equals $\phi$. The electrical charge transfer between the donor and the acceptor is due to the electric potential difference between the $A$ and $B$ compartments, and leads eventually to the potential equilibrium ($\phi_A = \phi_B$). Similarly, a concentration difference between the donor and acceptor compartments (measured with respect to an arbitrary value of $c_{ref} = 0$) leads to the diffusive mass transfer as long as $c_A = c_B$.

---

[2] This assumption will be removed in section 4.1.

The one-to-one similarity proceeds by setting the correspondence between the electrical capacitance $C$ and the mass volume $V$ and between the electrical resistance $R_e = \frac{\rho_e l}{a} = \frac{l}{\kappa_e a}$, and the mass resistance:

$$R = \frac{l}{Da}, \qquad (2.2)$$

which can be defined as the hindrance that molecule encounters passing through a medium of diffusivity $D$, length $l$ and enclosing surface $a$ (fig. 2).

The *release time T* of a MTS is defined as the time occurring to reach the concentration equilibrium. This is an asymptotic value and the estimation of $T$ depends on the accuracy required. On the other hand, in drug release process, the mass release is expressed through a Weibull function, written in the form (Weibull, 1951):

$$\frac{M(t)}{M_0} = \exp\left[-\left(\frac{t}{\tau}\right)\right] \qquad (2.3)$$

where $M(t)$ is the mass at time *t*, $M_0$ is the initial mass and the parameter $\tau$ (*characteristic time*) is associated with the time where approximately 63% of the drug is delivered[3]. The empirical exponential form of Eq.2.3 allows to fit a variety of experimental release data and is commonly used to obtain phenomenological insights into the intrinsic mechanisms governing the dissolution and drug release processes (Papadopoulou et al., 2006).

For a system of characteristic length $l$ and diffusivity $D$, the characteristic time is defined as $\tau = RV = \frac{l^2}{D}$ that corresponds to the analogue time constant[4] $\tau_e = R_e C$ for a RC circuit. By definition, $\tau$ measures the response of a body to transfer its mass and is linearly dependent on the volume and on the resistivity of the medium. The full correspondences between the mass, heat transfer parameters and their electric analogues are summarized in Table 1.

---

[3] In general the Weibull function is defined as a stretched exponential function $\frac{M(t)}{M_0} = \exp\left[-\left(\frac{t}{\tau}\right)^b\right]$ where the case b≠1 is usually associated with porosity and/or memory effects (Casault and Slater, 2009; Gomes-Filho et al., 2020).

[4] The counterpart time constant $\tau_e$ = RC is the time required to charge a capacitor, through the resistor R, from an initial 0 voltage to a 63% of the value of an applied DC voltage.

At this point we are in the position to make an important assumption from now on. In a typical MTS, it is common that one of the two compartment has a very low resistance to the mass transfer (or very high diffusivity $D$) compared with the other one. This implies that the transfer process is entirely dominated by the slowest diffusive medium that, from now on, is identified with the $B$ compartment. In other words, we assume that $A$ behaves as an "unresistive" compartment, i.e. $\tau_A = R_A V_A = \frac{l_A^2}{D_A} \approx 0$, and, as a consequence, the characteristic time of the whole transfer process identifies with $\tau_B$. By keeping this in mind, henceforth we set $D_B \equiv D$, $l_B \equiv l$ and $\tau_B = \tau$.

Tab. 1: Analogy of variables between EC and MTS. For convenience, in the last column the correspondent variables for the heat transfer are listed.

| *Transport analogies* | *Electrical (e)* | *Mass (m)* | *Heat (Q)* |
|---|---|---|---|
| Flux (current density) | $\vec{J_e}$ | $\vec{J_m}$ | $\vec{J_Q}$ |
| Characteristic variable | electric potential $\phi$ | mass concentration $c$ | temperature $\theta$ |
| Driving force | electric potential gradient $\nabla \phi = \frac{\Delta \phi}{\Delta l}$ | concentration gradient $\nabla c = \frac{\Delta c}{\Delta l}$ | temperature gradient $\nabla \theta = \frac{\Delta \theta}{\Delta l}$ |
| Proportionality factor | electrical conductivity $\kappa_e$ | diffusivity $D$ | thermal conductivity $\kappa_Q$ |
| *RC parameters* | | | |
| Voltage $U$ | $\Delta \phi$ | $\Delta c$ | $\Delta \theta$ |
| Current $I$ | $\dot{e} = j_e \ a$ | $\dot{m} = j_m \ a$ | $\dot{Q} = j_Q \ a$ |
| Resistance $R$ | $\frac{l}{\kappa_e a}$ | $\frac{l}{Da}$ | $\frac{l}{\kappa_Q a}$ |
| Capacitance $C$ | $e/\Delta\phi$ | $m/\Delta c = V$ | $Q/\Delta\theta$ |
| Characteristic time $\tau$ | $R_e C$ | $R V = l^2/D$ | $lQ/\kappa_Q a \Delta\theta$ |

### 3. Solution of the problem

In this section we will use the 1-to-1 analogy between MTS and RC circuit, based on the Kirchhoff's voltage (KVL) or current (KCL) laws respectively[5] (Nilsson and Riedel, 2010), to solve the MTS problem analytically as follows.

Let us consider two bodies, $A$ and $B$, of volumes $V_A$ and $V_B$, in a perfect contact, as a closed insulated system of total mass M, such that a variation of mass in the $B$ compartment yields a correspondent opposite variation of mass in $A$, with the total mass remaining constant at any time:

$$m_A(t) + m_B(t) = M \qquad (3.1)$$

For the following use, let us define the average concentration $\bar{c} = M/V$, with $V = V_A + V_B$ as the total volume. By recalling that $c_A(t) = {m_A(t)}/{V_A}$, the KVL for an RC loop (Fig.1) is given by $\phi_A(t) = \dot{e}_B(t)R_e + e_B(t)/C_B$, and its MTS analogue reads:

$$c_A(t) = m_A(t)/V_A = \dot{m}_B(t)R + m_B(t)/V_B, \qquad (3.2)$$

with the initial conditions[6]:

$$m_B(0) = m_B^0 = c_B^0 V_B \qquad (3.3)$$

By Eq.(3.1), a general solution of ODE's (3.2)-(3.3) is:

$$m_B(t) = V_B c_B(t); \qquad c_B(t) = \bar{c} + (c_B^0 - \bar{c})\exp\left(-\frac{t}{\tau}\right), \qquad (3.4)$$

---

[5] KVL law states that the sum of voltage in any loop is equal to zero ($\sum U = 0$), while KCL law expresses that the sum of currents flowing into a node is equal to the sum of currents flowing out of the node ($\sum I = 0$) (Nilsson and Riedel, 2010).
[6] An equivalent way is to consider the current as proportional to the potential difference between $A$ and $B$, yielding $I(t) \equiv \dot{m}_B(t) = V_B \dot{c}_B(t) = \frac{c_A(t) - c_B(t)}{R}$, which, after some algebra, recovers Eq.(3.2).

where: $\tau = RV_{eq}$ with $V_{eq} = V_A V_B/(V_A + V_B)$. This equivalent volume is computed as for the electrical analogue of two capacitors in series, i.e. $\frac{1}{V_{eq}} = \frac{1}{V_A} + \frac{1}{V_B}$.

The equilibrium quantities are: $\lim_{t\to\infty} c_B(t) = \bar{c}$ and $m_\infty = \lim_{t\to\infty} m_B(t) = \bar{c}\, V_B$. For the special case $c_B^0 = 0$, Eq.3.4 reduces to:

$$m_B(t) = V_B\, c_B(t) \qquad c_B(t) = \bar{c}\left[1 - \exp\left(-\frac{t}{\tau}\right)\right] \qquad (3.5)$$

For $c_B^0 > \bar{c}$ the mass transfers from $B$ to $A$ (*release*), while for $c_B^0 < \bar{c}$ one deals with $A \to B$ transfer (*absorption*). Results of the simulations of the concentration evolution at donor/acceptor compartments for different ratio $V_A:V_B$ ($V_A=20$, $c_B^0=0$, $M=20$, $\tau=1$) are shown in fig.3.

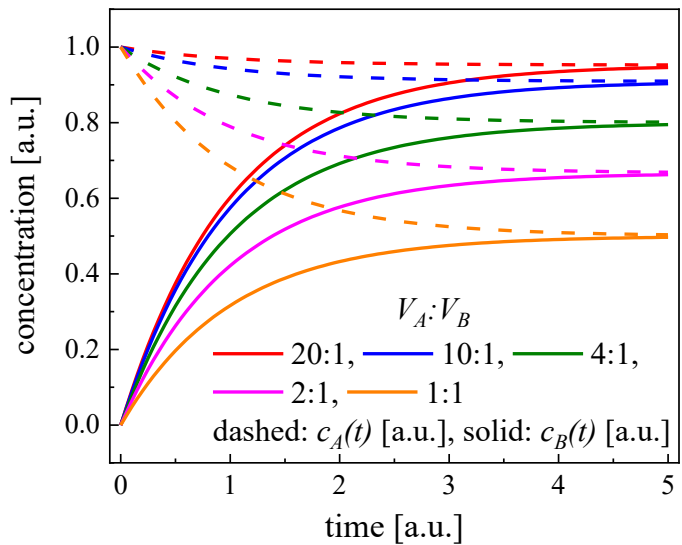

Fig.3. Concentration profiles (Eq.3.5) for $c_B(t)$ (solid lines) and $c_A(t)$ (dashed lines) for various ratios of $V_A:V_B$ ($V_A=20$, $c_B^0=0$, $M=20$, $\tau=1$).

### 3.1 Special case: $V_A \gg V_B$

In most situations, one of the compartments has a volume much larger than the other, as in the common case of a finite-size body placed in a *large* environment. To fix ideas, let us assume that $V_A \gg V_B$ such that $c_A$ can be considered as constant (the percentage variation of concentration/mass of $A$ is very small). Therefore $A$ can be modelled as a DC element or a capacitor with $C_A \gg C_B$. Then, the Kirchhoff's II law for MTS analogue can be written (cfr. Eqn (3.2)):

$$c_A = \dot{m}_B(t) R + m_B(t)/V_B, \quad (3.6)$$

$$m_B(0) = m_B^0 = c_B^0 V_B. \quad (3.7)$$

where $c_A$ is a constant.

The general solution of Eqs 3.6-3.7 is given by:

$$m_B(t) = V_B c_B(t) \; ; \quad c_B(t) = c_A + (c_B^0 - c_A) \exp\left(-\frac{t}{\tau}\right) \quad (3.8)$$

where , in this case,

$$\tau = R\, V_{eq} = R\, V_B \quad (3.9)$$

Eq.3.8 describes mass transfer for both cases: absorption for $c_A > c_B^0$, (Fig. 1a) and desorption for $c_A < c_B^0$, (Fig. 1b).

Note that the solution (3.8) is formally identical to Eq.3.5 with $\bar{c} = c_A$ and gives the asymptotic expressions: $\lim_{t\to\infty} c_B(t) = c_A$ and $m_\infty = \lim_{t\to\infty} m_B(t) = c_A V_B$. Simulations for $V_A = 20$, $V_B = 1$, $M = 20$, $\tau = 1$ were carried out. The concentration evolutions for $c_B(t)$ with different $c_B^0$ are shown in fig.4.

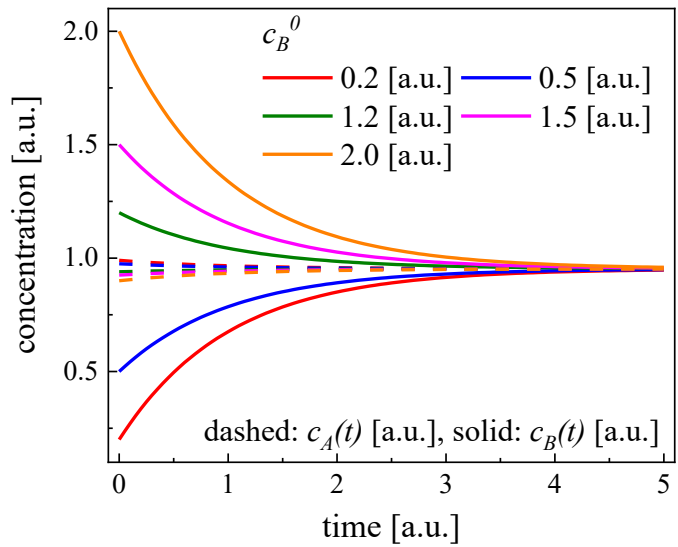

Fig.4. Concentration profiles (Eq.3.8) for: $c_A(t)$ (dashed lines) and $c_B(t)$ (solid lines) under various $c_B^0$ ($V_A=20$, $V_B=1$, $M=20$ and $\tau=1$).

Two special cases are noteworthy:

*Absorption with $c_B^0 = 0$*

Eq. 3.8 becomes

$$m_B(t) = V_B\, c_B(t); \qquad c_B(t) = c_A \left[1 - \exp\left(-\frac{t}{\tau}\right)\right]. \qquad (3.10)$$

*Release with $V_A \to \infty$, $c_A = 0$*

The depletion of the $B$ compartment is given by:

$$m_B(t) = V_B\, c_B(t); \qquad c_B(t) = c_B^0 \exp\left(-\frac{t}{\tau}\right) \qquad (3.11)$$

and the correspondent release profile is:

$$m_A(t) = m_B^0 \left[1 - \exp\left(-\frac{t}{\tau}\right)\right] \qquad (3.12)$$

Concentration curves for $c_B(t)$ (absorption case) under various initial concentrations $c_B^0$, and $c_A = 1$ and $\tau = 1$, are shown in fig.5.

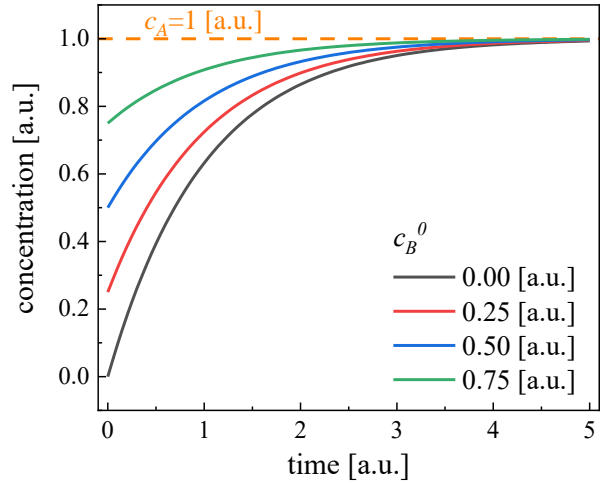

Fig.5. Absorption concentration profiles (Eq.3.10) under various initial concentrations $c_B^0$ and constant concentration at the source ($c_A=1$) and $\tau=1$.

The symmetric case of $V_A \ll V_B$ can be obtained in a similar way, by swapping the role of $A$ and $B$ compartments.

## 4. Mass diffusion through shells

Let us extend the results of the previous section to the case when the body B is separated from A by one or multiple thin shells or membranes (Fig.6a) that provide a protective cover and a mechanical barrier (*coating*) for the mass diffusion (Kaoui et al., 2018).

First, let us consider the body $B$ (donor or acceptor) of characteristic length $l$ and resistance $R$ faced with a single thin coating, shell or membrane $S$, of thickness $l_S$, diffusivity $D_S$ and resistance $R_S$, and immersed in a very large or infinite medium A ($V_A \gg V_B$). In the general case, one has to enrich the EC model by adopting another capacitor mimicking the shell presence (Fig.6b). To find the concentrations evolutions at the $B$ and $S$ compartments for the $A \to B$ transfer case, it is convenient to switch to the current representation (KCL), yielding the set of ODE's:

$$\dot{m}_S(t) = V_S \dot{c}_S(t) = \frac{c_A - c_S(t)}{R_S} - \frac{c_S(t) - c_B(t)}{R}, \tag{4.1a}$$

$$\dot{m}_B(t) = V_B \dot{c}_B(t) = \frac{c_S(t) - c_B(t)}{R}. \tag{4.1b}$$

To find the exact solution for $c_B(t)$ one has to solve a second-order differential equation. The solution does not provide a simple characteristic time formula, as obtained earlier for a perfect contact donor/acceptor systems. In particular, the insertion of an additional shell compartment yields a multi-layer transfer system, where the concentration evolution at the intermediate layer (shell) is influenced by the concentrations at the neighbouring compartments (see eqn (4.1a)).

In the general $A \to B$ multi-layer case (layers 1,2,…n, with $n = B$), Eq.4.1 can be generalized to a system of ODE's:

$$\dot{m}_1(t) = V_1 \dot{c}_1(t) = \frac{c_A - c_1(t)}{R_1} - \frac{c_1(t) - c_2(t)}{R_2},$$

$$\dot{m}_2(t) = V_2 \dot{c}_2(t) = \frac{c_1(t) - c_2(t)}{R_2} - \frac{c_2(t) - c_3(t)}{R_3}, \tag{4.2}$$

$$\ldots\ldots\ldots$$

$$\dot{m}_n(t) = V_n \dot{c}_n(t) = \frac{c_{n-1}(t) - c_n(t)}{R_n}.$$

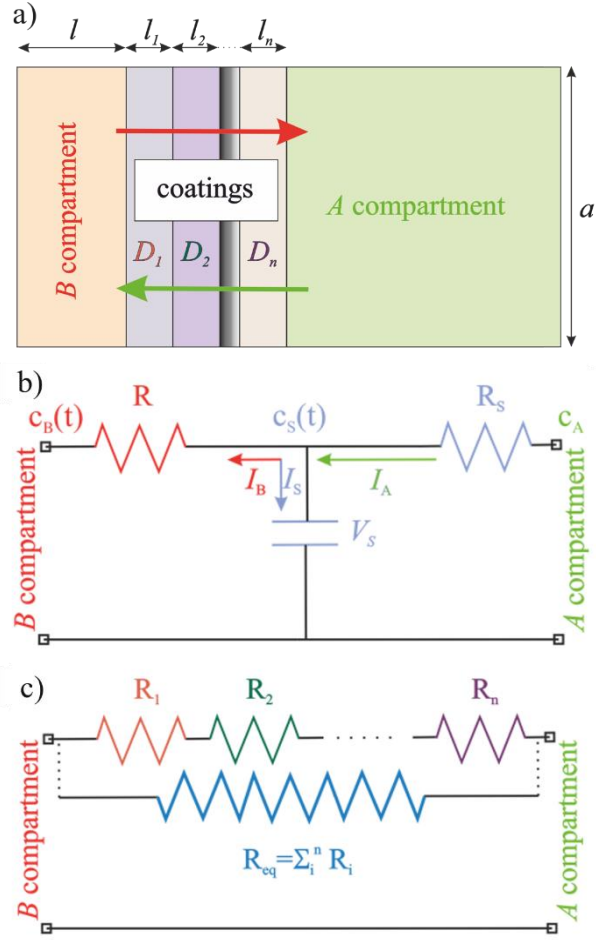

Fig 6. Schematic illustration of the multi-layer mass diffusion (a), the EC circuit for an absorbing single-shell case with a slowly-equilibration conditions (b), and the corresponding multi-resistance electrical circuit under fast-equilibration conditions (c).

*4.1 Fast-equilibration conditions*

If the shell undergoes fast equilibration ($\tau_S \ll \tau$ or $\frac{l_S^2}{D_S} \ll \frac{l^2}{D}$), the shell capacitor can be omitted, and the equivalent resistance of the system corresponds to the resistors connected in series, i.e: $R_{eq} = R + R_S = \frac{l}{Da} + \frac{l_S}{D_S a} = \frac{1}{a}\left(\frac{l}{D} + \frac{l_S}{D_S}\right)$.

Then, the characteristic time of the system (see Eq. 3.9) is given by:

$$\tau = R_{eq} V_B = \frac{l}{Da} al + \frac{l_S}{D_S a} al = l\left(\frac{l}{D} + \frac{l_S}{D_S}\right). \qquad (4.3)$$

Similarly, if $B$ is surrounded by $n$ homogenous thin layers of thickness $l_i$, and diffusivity $D_i$, (i=1,2,..,n, Fig.6c), the equivalent resistance experienced by the diffusing mass with $l \equiv l_0$ is: $R_{eq} = \sum_{i=0}^{n} R_i$ and Eq.4.3 extends to:

$$\tau = l_0 \sum_{i=0}^{n} \frac{l_i}{D_i} \tag{4.4}$$

*4.2 General equilibration conditions – approximated solution*

In the case of the mass transfer through a non-zero capacitance shell, under: $c_A = 1$, $c_B^0 = 0$ and $c_S^0 = 0$, the approximated characteristic time can be given as:

$$\tau = R_S V_S + (R + R_S) V_B = \frac{l_S}{D_S a} a l_S + \frac{1}{a}\left(\frac{l}{D} + \frac{l_S}{D_S}\right) al = \frac{l_S^2}{D_S} + l\left(\frac{l}{D} + \frac{l_S}{D_S}\right), \tag{4.5}$$

which corresponds to the sum of the shell characteristic time $\tau_S = \frac{l_S^2}{D_S}$ and the single donor/acceptor $\tau$ given by Eq.4.3.

Two cases are considered as examples: i) the shell and the $B$ compartment have similar characteristics ($R = R_S = 1$, $V_B = V_S = 1$), and ii) the shell provides a rate limiting compartment for the mass transfer ($R = 1$, $R_S = 100$, $V_B = 100$, $V_S = 1$). Fig.7 shows the correspondent concentrations.

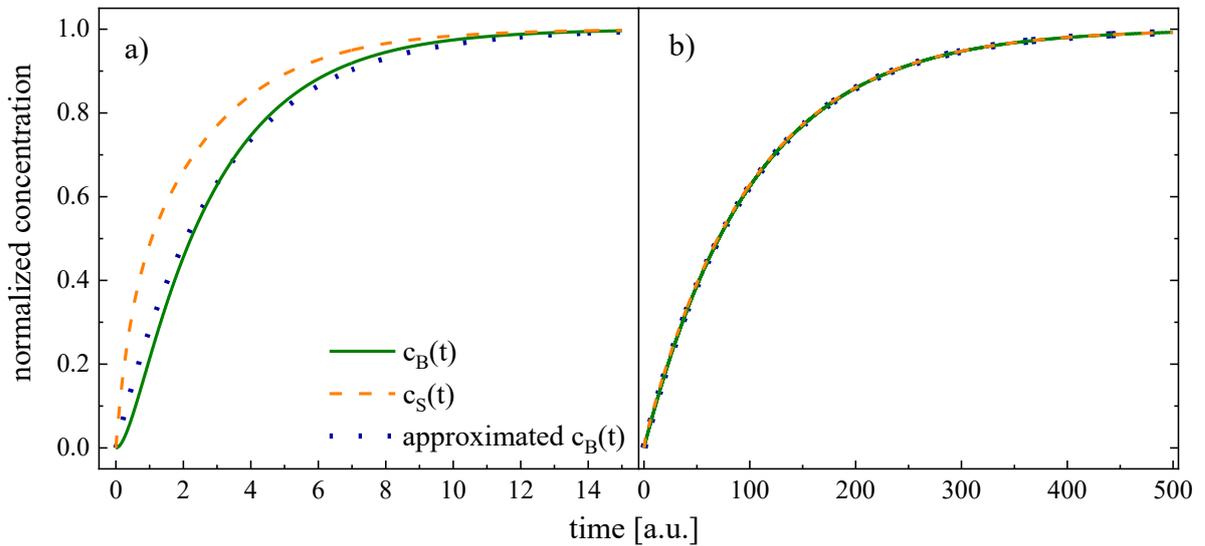

Fig.7 The shell and acceptor compartment absorption curves $c_B$, $c_S$ with $c_A = 1$, $c_B^0 = 0$ and $c_S^0 = 0$. Fig. a) shows the concentration evolution when $S-B$ have similar thickness and diffusivity, while b) show the concentration evolutions in the presence of thin, resistive shell ($c_B(t)$, $c_S(t)$ and the approximated $c_B(t)$ (Eq.4.5, blue, dotted), all are overlapping).

## 5. Two-phase diffusion

Let us generalize the problem described in section 2 in another direction, by considering the $B$ compartment constituted by two distinct phases, i.e. a network of permeable channels (volume I, phase I) in a solid matrix (volume II, phase II) (i.e. a porous material) or a composite medium having a microstructure made of two differently permeable materials (Fig.8). The mass diffuses in $B$ over these two phases: with resistance $R_I$ in phase $I$ and with resistance $R_{II}$ in phase $II$. By incorporating homogenization procedure (i.e. by averaging the system characteristics over a certain representative volume), the equivalent resistance in $B$ under no-intermixing conditions corresponds to the two resistors connected in parallel. Assuming that the cross-section area of the source $a$ is split into $I$ and $II$ contributions ($a = a_I + a_{II}$), the equivalent resistance of $B$ can be written as:

$$R_{eq} = \frac{R_I R_{II}}{(R_I + R_{II})} = \frac{l^2}{D_I a_I D_{II} a_{II}} \frac{1}{l} \left( \frac{D_I a_I + D_{II} a_{II}}{D_I a_I D_{II} a_{II}} \right)^{-1} = \frac{l}{D_I a_I + D_{II} a_{II}}. \qquad (5.1)$$

By denoting the porosity by $p = V_I/V_B = a_I/a$, the sub-areas are given by: $a_I = pa$ and $a_{II} = (1-p)a$. Then, Eq.5.1 can be written in terms of the total cross-section area as:

$$R_{eq} = \frac{l}{a\widetilde{D}}. \qquad (5.2)$$

with $\widetilde{D} = pD_I + (1-p)D_{II}$ as a volume averaged diffusivity.

In case of a very large or infinite medium $(V_A \gg V_B)$, the concentration evolution under the condition $c_B^0 = 0$ obeys the exponential kinetics (as in Eq.3.8) with the characteristic time $\tau = l^2/\widetilde{D}$. In the special case $D_I = D_{II}$, Eq.5.2 yields $R_{eq} = \frac{l}{D_I a} = R$, that recovers the previous result for the homogeneous case (Eq.2.2).

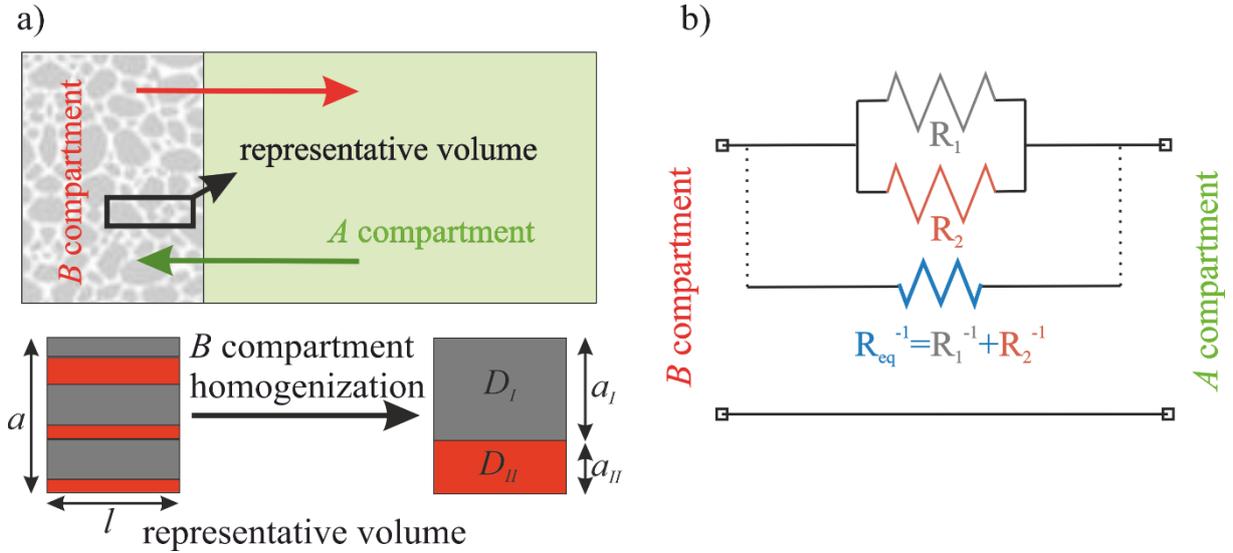

Fig 8: Schematic illustration of the two-phase mass diffusion (a) and the correspondent multi-resistance electrical circuit (b).

## 6. Validation

We now validate our model predictions against analytical and numerical data on drug kinetics and on the method efficacy, thus providing confidence on the current modeling approach.

### 6.1 Recovering classical solutions

In his comprehensive book on diffusion problems (Crank, 1975), Crank provides the exact solutions for the diffusion-driven mass problems from systems of various geometries. In particular, three examples of drug-loaded bodies of various geometry releasing in A are considered:

a) plane sheet of thickness $2d$ with constant surface concentration ([2], Eq.4.18):

$$\frac{m_A(t)}{m_\infty} = 1 - \sum_{j=0}^{\infty} \frac{8}{(2j+1)^2 \pi^2} \exp\left[\frac{-D(2j+1)^2 \pi^2 t}{(2d)^2}\right]; \quad (6.1a)$$

b) cylinder of radius $r$ with constant surface concentration ([2], Eq.5.23):

$$\frac{m_A(t)}{m_\infty} = 1 - \sum_{j=1}^{\infty} \frac{4}{r^2 \beta_j^2} \exp[-D\beta_j^2 t], \quad (6.1b)$$

where $\beta_j$ are the roots of $J_0(r\,x) = 0$, with $J_0$ the Bessel function of the first kind and zero order;

c) sphere of radius $r$ with constant surface concentration ([2], Eq.6.20):

$$\frac{m_A(t)}{m_\infty} = 1 - \frac{6}{\pi^2}\sum_{j=1}^{\infty}\frac{1}{j^2}\exp[\frac{-Dj^2\pi^2 t}{r^2}]. \qquad (6.1c)$$

To recover these solutions with our electrical analogue approach, we generalize Eq.3.12 with a correction as:

$$\frac{m_A(t)}{m_\infty} = 1 - \sum_{i=1}^{\infty} F_i \exp\left[-\alpha_i \frac{Dt}{l^2}\right] = 1 - \sum_{i=1}^{\infty} F_i \exp\left[-\alpha_i \frac{t}{\tau}\right], \qquad (6.2)$$

where $F_i$ and $\alpha_i$ are the geometry-correction coefficients. A direct comparison between (6.2) and (6.1) leads to the expression for $F_i$ and $\alpha_i$ for any specific geometry. For example, in the case c) we have:

$$\tau = \frac{r^2}{D}; \quad \alpha_i = i^2\pi^2; \quad F_i = \frac{6}{i^2\pi^2} = \frac{6}{\alpha_i}; \qquad (6.3)$$

Computationally, the series presented in eqs. 6.1-6.2 are truncated to some finite $n$ that guarantees convergence in the specific geometry. Similar procedures can be found in the literature concerning analysis of pharmacologically-relevant systems, where the absorption/release processes are characterized by a certain order kinetics (Zhang et al., 2010). To find $F$ and $\alpha$ parameters for the mass release problems governed by mono- and bi-exponential kinetics, a best-fit procedure of Eq.6.2 over Eqs 6.1a-c with $D = 1$ and $l = d = r = 1$ has been applied. For the models provided by Eqs 6.1, the summation over $j$s has been limited to n= 5 terms. In the case of Eq.6.2, only two first terms are considered; for the $n$=2, the additional condition $F_1 + F_2 = 1$ has been imposed. The results of the best-fit procedure are presented in Fig.9, while the optimal parameters are given in Table 2.

Tab.2 The best-fit parameters of Eq.6.2 to the diffusion-driven mass release models for various system geometries.

| Eq.6.2 | geometry | | |
|---|---|---|---|
| | plane sheet | cylinder | sphere |
| $n$=1 | | | |
| $F_1$ | 0.86±0.01 | 0.83±0.01 | 0.82±0.01 |
| $\alpha_1$ | 2.62±0.02 | 6.81±0.15 | 12.9±0.35 |
| $n$=2 | | | |
| $F_1$ | 0.82±0.01 | 0.72±0.02 | 0.64±0.04 |
| $\alpha_1$ | 2.49±0.01 | 5.92±0.13 | 10.2±0.5 |
| $F_2$ | 0.18±0.01 | 0.28±0.02 | 0.36±0.04 |
| $\alpha_2$ | 47.9±2.71 | 70.4±11.1 | 90.8±23.6 |

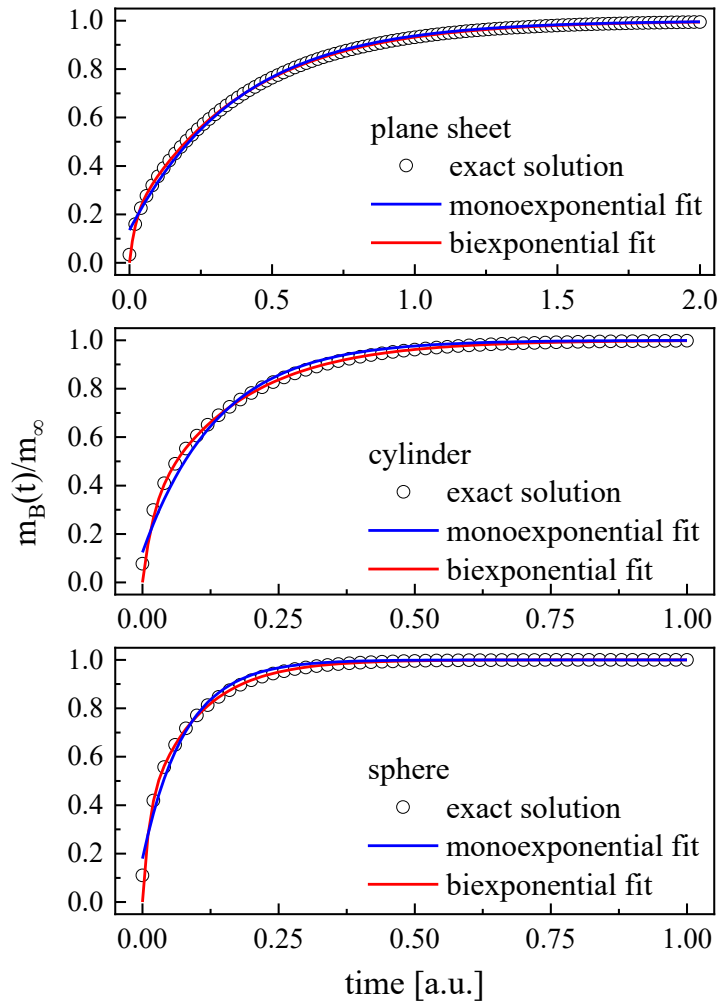

Fig.9 Best fits of the EC model (mono- (*n*=1) and biexponential (*n*=2) fitting functions) to the predictions of the diffusion-driven mass release models (exact solutions provided by Eqs 3.3) for various system geometries. The characteristic lengths and the diffusion coefficients have been set to unity.

**6.2 Drug diffusion through a two-phase membrane**

In their work, Bodzenta *et al.* analyzed drug transport into a composite dodecanol-collodion membrane (i.e. plane sheet geometry) by means of Fick's II law diffusion model (Bodzenta et al., 2006). The experimental part of the work relied on the time-dependent FTIR-ATR spectra acquisition of the membrane (of thickness $l = 25 \pm 5$ μm) in contact with a semisolid formulation containing dithranol as a drug. The normalized magnitude of the ATR signal corresponding to the dithranol amount in the membrane is provided in Fig.10 (blue circles). As shown by the authors, the diffusion-based model was in good agreement with the experimental data, yielding a drug diffusion coefficient of D= $(2.1 \pm 0.4) \cdot 10^{-10}$ cm$^2$s$^{-1}$.

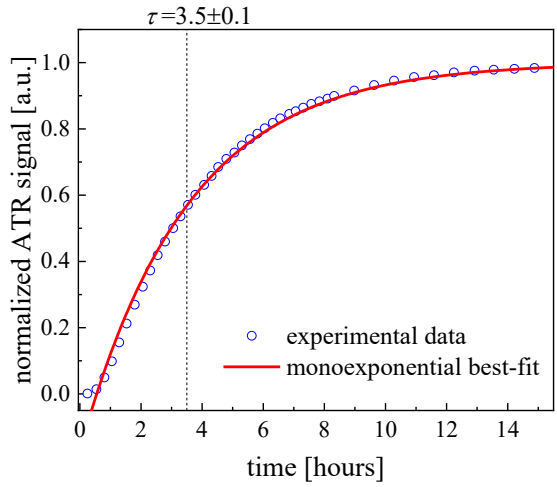

Fig.10 Experimental data (blue circles) for dithranol diffusion into collodion-based membrane studied by means of FTIR-ATR (from (Bodzenta et al., 2006)) and the mono-exponential EC model best-fit (Eq. 6.3, red curve) with $c_B^0 = 0$.

The same dataset has been analysed by means of the EC model given by Eq.6.2 (the homogeneous membrane absorption with zero-initial drug concentration), as the drug source has been considered as constant during the whole experiment. The best-fit procedure applied to the data has revealed the characteristic time of the process $\tau = 3.5 \pm 0.1$ hours, while the computed goodness-of-fit parameter, the squared correlation coefficient, has been found to be 0.99. By keeping in mind the measured membrane thickness $l$, the magnitude of the characteristic time $\tau$, and considering the correction coefficient $\alpha_1$ (see Tab.2), the EC diffusion coefficient, provided by $D = l^2/\alpha_1\tau$, equals to $(1.9 \pm 0.6) \cdot 10^{-10}$ cm$^2$s$^{-1}$, which is comparable with the value reported in (Bodzenta et al., 2006).

By referring to the membrane synthesis procedure described in (Neubert et al., 1991), it is possible to roughly estimate the *dry* collodion/dodecanol membrane compounds mass ratio to be 2/3. On the other hand it is known that the drug penetrates only through dodecanol-filled pores of the membrane, while the collodion matrix provides an impermeable phase for drug diffusion (Rochowski et al., 2020). As such, the membrane can be considered as porous (as such, $D \equiv \widetilde{D}$ for the diffusion coefficient examined earlier), with the porosity of $p = 60\%$, and the EC membrane resistance is given by Eq.5.2. Finally, the drug diffusion coefficient through the membrane permeable structures, $D_p$, can be estimated as $D_p = \widetilde{D}/p \approx (3.2 \pm 1.0) \cdot 10^{-10}$ cm²s⁻¹.

## 6.3 Diffusion through multi-layer systems

The EC model has been validated for different two-layer systems. First, we compare the exact solutions for the homogeneous case (i.e. release from a plane sheet and a sphere - Eqs 6.2a and 6.2c) with the EC model (Eq.4.1 – general equilibration conditions). This has been simulated over a *heterogeneous-like system*, comprising of two subsystems of equal diffusivities of $D$ =3·10⁻¹⁰ m²s⁻¹. The EC simulations are performed for the following conditions: $c_B^0 = c_S^0 = 1$ and $c_A = 0$. In particular, the overall characteristic length of an object (plane sheet/sphere of $l_o = r_o = 10^{-3}$ m, see fig. 11) is split into two characteristic lengths at various ratios (9:1, 1:1, 1:9). The inner $B$ compartmental parameters are: $R = \alpha \frac{l}{Da}$ and $V = al$, whilst the outer body ($S$ shell) characteristics are set as: $R_s = \alpha \frac{l_s}{Da}$ and $V = al_s$, with the $\alpha$ factor related to the planar and spherical geometry (see table 2 for n=1 case). The results of simulations are shown in Fig.12. It appears that EC model predictions are in good agreement with the results of the exact solutions, especially when $l_s/l$ ratio is far from unity.

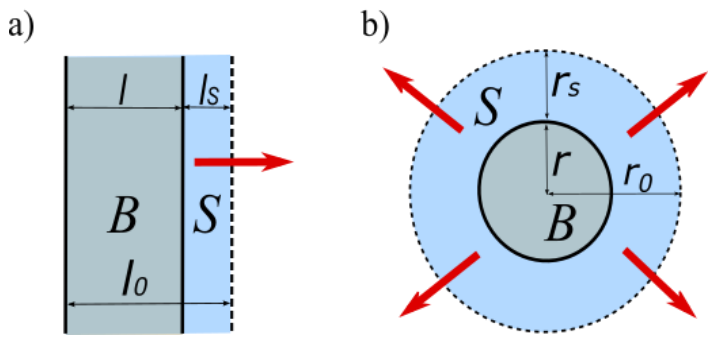

Fig. 11 The planar (a) and spherical (b) two-layer (B-S) systems used in the simulations.

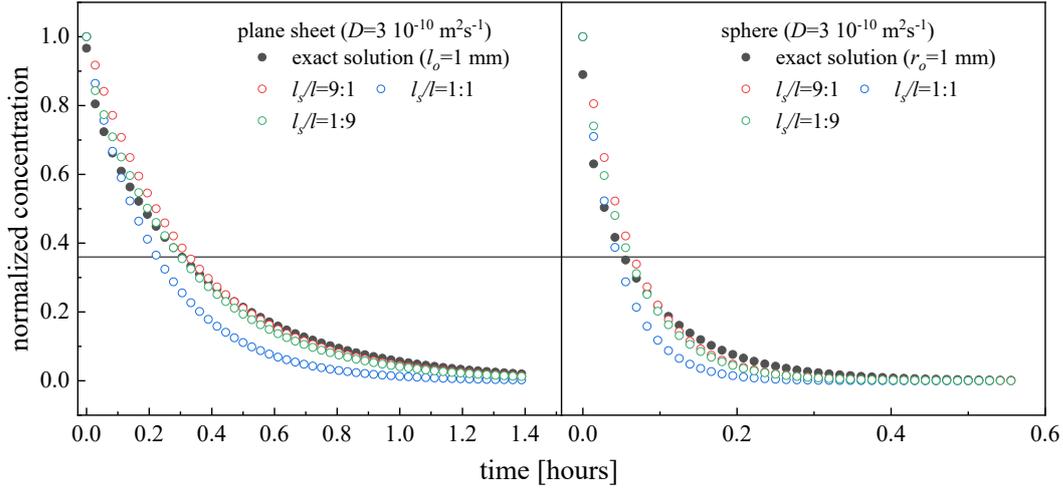

Fig.12 Comparison of the release profiles from a *heterogeneous-like* systems (plane sheet and sphere) of characteristic length $l=r=1\cdot 10^{-3}$ m and diffusivity of $D=3\cdot 10^{-10}$ m$^2$s$^{-1}$ into an infinite (c=0) medium – exact solutions vs EC multilayer approach predictions (Eq.4.5). The horizontal line marks the concentration value related to the release characteristic time.

It should be pointed out, that for *heterogeneous-like* systems comprising of thin/thick shells (the $l_S/l=1/9$ and 9) the overall release is dominated by one compartment only (quasi-homogeneous case), while for the $l_S/l=1$ one deals with a considerable mismatch between the shell EC ($V \propto l_S$) and actual ($V \propto l_o^3 - l^3$) volumes (fig. 12).

A second test involved release simulations from two-compartmental spherical capsule of radius $r_o$=1.7·10$^{-3}$ m, comprising of a core and a shell of diffusivities $D=3 \cdot 10^{-10}$ m$^2$s$^{-1}$ and $D_S = 2 \cdot 10^{-11}$ m$^2$s$^{-1}$ respectively, into an infinite environment ($c_A = 0$). Two cases were considered: 1) $c_B^0 = c_S^0 = 1$ and 2) $c_B^0 = 1$ and $c_S^0 = 0$, with distinct characteristic lengths: $l=r=1.3\cdot 10^{-3}$ m and $l_S=r_S=0.4\cdot 10^{-3}$m $\left(\frac{l_S}{l} = 0.3, \frac{D_S}{D} = 0.06\right)$, but comparable volumes: $V$=0.92·10$^{-8}$ m and $V_S$=1.14·10$^{-8}$ m (fig. 11 right). Due to the identified simple EC approach issues, the release model characteristics are given by: $R = \alpha \frac{l}{Dl^2}$ and $V = l^3$ for the core, and $R_S = \alpha \left(\frac{l_o}{D_S l_o^2} - \frac{l}{D l^2}\right) = \frac{\alpha}{D_S l_o} - R$ and $V_S = l_o^3 - l^3 = V_o - V$ for the shell compartment. As such, the shell is considered as a homogeneous sphere with a simple correction factor for the resistance of the core. Results of the simulation are shown in Fig.13b.

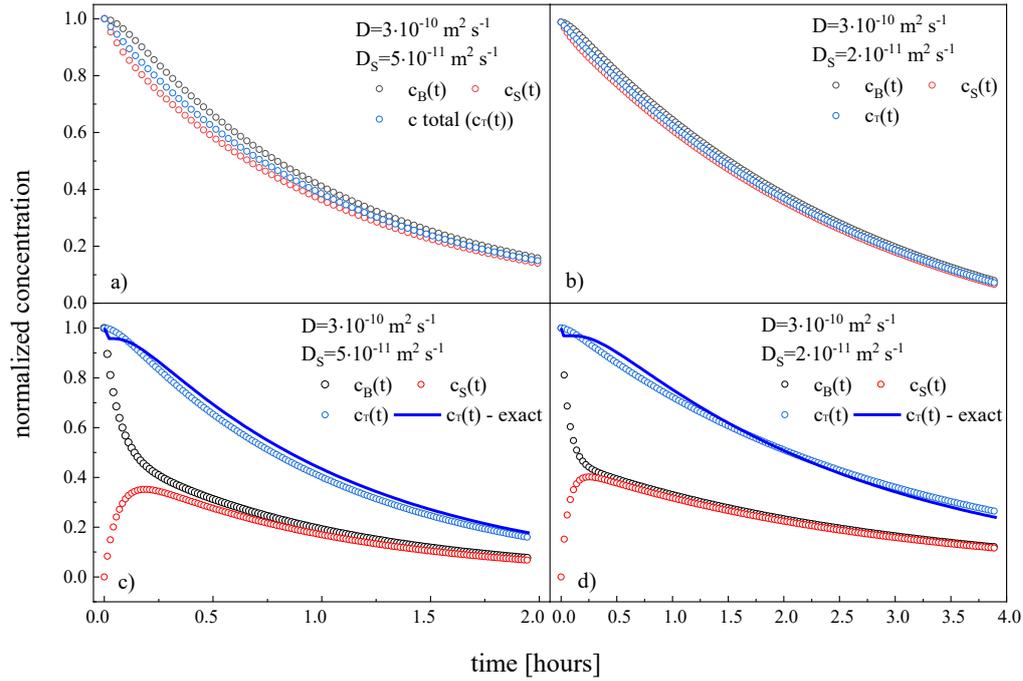

Fig.13 Release profiles from heterogeneous spherical capsules of $l_o=1.7 \cdot 10^{-3}$ m, involving a core ($l=1.3 \cdot 10^{-3}$ m) and a single shell ($l_s=0.4 \cdot 10^{-3}$ m) of distinct diffusivities, into an infinite medium ($c_A = 0$); a) and b) - $c_B^0 = 1$, $c_S^0 = 1$, c) and d) - $c_B^0 = 1$, $c_S^0 = 0$. Solid lines represent corresponding release profiles predicted by the model presented in (Kaoui et al., 2018).

The release data for the $c_B^0 = 1$ and $c_S^0 = 0$ case (Fig.13c) were confronted against analytical results obtained by means of the release model by Kaoui *et al.* (Kaoui et al., 2018). The authors considered a Fickian transport through a multi-layered system under various boundary conditions. For our purposes, the simulations are carried out with perfect contact conditions, implying continuity of fluxes and concentrations on each interface.

The release curves obtained by means of the exact model and EC approach revealed good agreement. In particular, the characteristic times predicted by the two models are similar: $\tau_{exact}$=1.20 h vs $\tau_{EC}$=1.11 h for the case a) and c) in Fig.13, and $\tau_{exact}$=2.86 h vs $\tau_{EC}$=2.85 h for the case b) and d).

## 7. Conclusions

Predicting the release performance of a mass transfer system is an important challenge in pharmaceutics and biomedical science. In this paper, inspired by drug delivery systems, we consider a multi-layer diffusion model of drug release from a composite body into an external surrounding medium. Based on this model, we present a simple approach for estimating the release time, i.e. the time required for the system to reach the equilibrium.

In some cases, in the absence of direct experiments or when is too complicated to compute an analytical or approximated solution, simple performance indicators can represent the main transport mechanisms and the most important components of the transfer process. In most circumstances, for example, rather than the fully distributed solution, the time required to reach a steady state drug concentration solely determines the effectiveness of the delivery system. Actually, the concept of characteristic time is crucial to pharmaceutical developers who want to create controlled- release devices able to deliver drugs at a desired rate. Without an explicit solution of the diffusion problem, the estimation of the *time constant* is of great importance in the design of delivery systems, because it allows product manufacturers to adjust certain properties to ensure a precise release within a determined time. The proposed methodology, which accounts for the relevant geometrical and physical parameters, has shown to capture the drug kinetics and provides a simple tool to measure drug delivery system performance.